\documentclass[aps,prb,twocolumn,superscriptaddress,%
preprintnumbers,amsmath,amssymb,floatfix]{revtex4}%

\usepackage{graphicx}
\usepackage{dcolumn}
\usepackage{bm}
\usepackage{amsmath}
\usepackage{amssymb}
\usepackage{revsymb}
 
\begin{document}

\affiliation{Department of Physics and Astronomy, Georgia State
University, Atlanta, Georgia 30303, USA}
\affiliation{Department of Chemistry, MIT, Cambridge, MA 02139, USA}
\affiliation{On sabbatical leave at Ecole Sup\'erieure de Physique et de Chimie Industrielle de la Ville de Paris, 10, rue Vauquelin, 75231 Paris, CEDEX 05, France}

\title{Nanoconcentration of Terahertz Radiation in Plasmonic Waveguides}

\author{Anastasia Rusina}
\affiliation{Department of Physics and Astronomy, Georgia State
University, Atlanta, Georgia 30303, USA}
\author{Maxim Durach}
\affiliation{Department of Physics and Astronomy, Georgia State
University, Atlanta, Georgia 30303, USA}
\author{Keith A. Nelson}
\affiliation{Department of Chemistry, MIT, Cambridge, MA 02139, USA}
\author{Mark I. Stockman}
\affiliation{Department of Physics and Astronomy, Georgia State
University, Atlanta, Georgia 30303, USA}
\affiliation{On sabbatical leave at Ecole Sup\'erieure de Physique et de Chimie Industrielle de la Ville de Paris, 10, rue Vauquelin, 75231 Paris, CEDEX 05, France}
\email{mstockman@gsu.edu}
\homepage{http://www.phy-astr.gsu.edu/stockman}

\date{\today}

\begin{abstract}
 
Recent years have seen an explosive research and development
of nanoplasmonics in the visible and near-infrared (near-ir) frequency
regions.%
\cite{Novotny_Hecht_2006_Principles_of_Nanooptics}
One of the most fundamental effects in nanoplasmonics is nano-concentration
of optical energy.
Plasmonic nanofocusing has been predicted
\cite{Phys_Rev_Lett_93_2004_Tapered_Plasmonic_Waveguides} and
experimentally achieved.%
\cite{Verhagen_Kuipers_Polman_Nano_Lett_2006_Tapered_Plasmonic_Waveguide,
Ropers_et_al_Nano_Lett_2007_Nanoconcentration_on_Cone,%
Verhagen_Polman_Kuipers_Opt_Express_2008_Adiabatic_Nanofocusing}
Nanoconcentration of optical energy at nanoplasmonic probes made
possible optical ultramicroscopy with nanometer-scale resolution%
\cite{Novotny_Stranick_ARPC_57_303_2006_NSOM,%
Lewis_et_al_Nature_Biotech_2003_SNOMs,%
Sandoghdar_et_al_Nano_Lett_2007_Fluorescence_Lifetime_Modification}
and ultrasensitive Raman spectroscopy.%
\cite{DiFabrizio_et_al_Nano_Lett_2008_Photonic_Nanoplasmonic_Device}
It will be very beneficial for the fundamental science, engineering,
environmental, and defense applications to be able to nano-concentrate
terahertz radiation (frequency $1-10$ THz or vacuum
wavelength $\lambda_0=300-30~\mathrm{\mu m}$). This will allow for
the nanoscale spatial resolution for THz imaging%
\cite{Mittleman_et_al_Rep_Progr_Phys_2007_THz_Imaging}
and introduce the THz spectroscopy on the nanoscale, 
taking full advantage of the rich
THz spectra and submicron to nanoscale structures of many engineering,
physical, and biological objects of wide interest: electronic components
(integrated circuits, etc.), bacteria, their spores, viruses,
macromolecules, carbon clusters and nanotubes, etc.
In this Letter we establish the principal limits for the
nanoconcentration of the THz radiation in metal/dielectric waveguides
and determine their optimum shapes required for this nanoconcentration
We predict that the adiabatic compression of THz radiation
from the initial spot size of $R_0\sim \lambda_0$ to the final size 
of $R=100-250$ nm can be achieved with the THz radiation intensity
increased by a factor of $\times10$ to $\times 250$. 
This THz energy nanoconcentration will not only 
improve the spatial resolution and increase the signal/noise ratio for
the THz imaging and spectroscopy, but in combination with the recently
developed sources of powerful THz pulses
\cite{Hebling_et_al_high_power_THz_sources_2008}
will allow the observation of nonlinear THz effects and a carrying out a variety of nonlinear spectroscopies (such as two-dimensional
spectroscopy), which are highly informative. This will find a wide
spectrum of applications in science, engineering, biomedical research,
environmental monitoring, and defense. 

\end{abstract}

\maketitle

There are existing approaches to deep subwavelength THz
imaging and probing based on sharp tips irradiated by a THz source,%
\cite{Chen_Kersting_Cho_APL_2003_THz_Imaging_with_100_nm_resolution}
adiabatically-tapered metal-dielectric waveguides%
\cite{Klein_et_al_JAP_2005_Metal_Dielectric_Antenna_for_THz_SNOM}
similar to optical adiabatic concentrators,%
\cite{Phys_Rev_Lett_93_2004_Tapered_Plasmonic_Waveguides,%
Verhagen_Kuipers_Polman_Nano_Lett_2006_Tapered_Plasmonic_Waveguide,%
Ropers_et_al_Nano_Lett_2007_Nanoconcentration_on_Cone,%
Verhagen_Polman_Kuipers_Opt_Express_2008_Adiabatic_Nanofocusing}
and nonlinear microscopic THz sources.%
\cite{Lecaque_Gresillon_Boccara_Opt_Expr_2008_THz_Emission_Microscopy}
For the development of the THz nanotechnology, it is extremely important
to understand spatial limits to which the THz radiation energy can be
concentrated (nanofocused).

A major challenge for the nanoconcentration of the electromagnetic
energy in the THz region is the large  radiation wavelength in vacuum or
conventional dielectrics,
$\lambda_0=30~\mathrm{\mu m}-300~\mathrm{\mu m}$, where the THz radiation
can only be focused to the relatively very large regions of size $\sim
\lambda_0/2$. The developed field of optical energy concentration, which is based
on surface plasmon polaritons (SPPs),
suggests that one of the ways to solving
this problem is to employ the surface electromagnetic
waves (SEWs). In the far infrared (ir), the dielectric 
permittivity of metals has large
imaginary part which dominates over its negative real part.%
\cite{Ordal_fir_metal_dielectric_permittivities_ApplOpt_1983}
This implies that SEWs propagating along a metal-dielectric
flat interface in this frequency range, known as
Sommerfeld-Zenneck  waves,%
\cite{Sommerfeld_Ann_Phys_Chem_1899, Zenneck_Ann_Phys_1907} are weakly bound
to the surface\cite{Saxler_TSP_PRB_2004}
and can hardly be used for the confinement of
THz radiation.

It has been suggested that periodically perforating flat surfaces
of ideal metals
with grooves or holes leads to the appearance of SEWs, which mimic
(``spoof'') SPPs to be stronger bound to the surfaces%
\cite{Mills_Maradudin_roughness_PRB_1989,
pendry_spoof, Garcia_Vidal_Planar_Spoof_JOA_2005}
permitting a better control over the THz fields. It has been
predicted that SPPs on an array of parallel
grooves cut on the surface of a perfect conductor
wire can be localized by adiabatic deepening of the grooves.%
\cite{Maier_Andrews_Martin-Moreno_Garcia-Vidal_PRL_%
2006_THz_SPPs_in_Periodically_Corrugated_Wires}
At a point, where grooves are approximately a
quarter of wavelength of light, the highest concentration is
achieved to be on the order of
tens of micrometers. This method restricts the localization
point to a particular frequency, making
the concentration very narrow-band. Also, the depth of a groove should
be $\approx \lambda_0/4$, i.e., in the tens to hundred micron range,
which precludes completely nanoscale devices.

It is well known from microwave technology that the ideal-metal
waveguides with smooth surfaces support TEM waves,
where the electric field lines are either
infinitely extended or terminate at the metal surfaces normally to them.
The latter case requires the waveguide cross-section topology
to be more than  single-connected; an example may be a coaxial
waveguide (``coax''). Such waveguides possess are very wide-band in
frequency. The THz waveguides can be adiabatically tapered
to concentrate energy. The idea of adiabatic energy concentration
comes from ultramicroscopy%
\cite{Keilmann_nonadiabatic_coax_FIR_microscopy_1995,%
Keilmann_et_al_Micron_1996_Coaxial_Tip,%
Keilman_et_al_Opt_Commun_2006_Coaxial_Tips}
and nanoplasmonics,%
\cite{Phys_Rev_Lett_93_2004_Tapered_Plasmonic_Waveguides}
where it has been developed both theoretically and experimentally%
\cite{Verhagen_Kuipers_Polman_Nano_Lett_2006_Tapered_Plasmonic_Waveguide,%
Ropers_et_al_Nano_Lett_2007_Nanoconcentration_on_Cone,%
Verhagen_Polman_Kuipers_Opt_Express_2008_Adiabatic_Nanofocusing}
and used in ultrasensitive surface enhanced Raman spectroscopy.%
\cite{DiFabrizio_et_al_Nano_Lett_2008_Photonic_Nanoplasmonic_Device}
Employing these ideas of the adiabatic concentration
and using a tapered metal-dielectric waveguide,
the THz spatial resolution achieved is $\sim
20~\mathrm{\mu m}$ across the entire THz spectrum.%
\cite{Klein_et_al_JAP_2005_Metal_Dielectric_Antenna_for_THz_SNOM}

In this Letter, for the first time, we establish the fundamental limits
and find the principles of designing the
optimum and efficient metal/dielectric waveguides suitable for
the THz nanofocusing.  The specific examples are for the wide-band
concentrators: a plasmonic metal wedge cavity and tapered coax
waveguides, which are terminated by funnel-type adiabatic tapers.
Such nano-concentrators along with the advent
of high-power sources%
\cite{Hebling_et_al_high_power_THz_sources_2008,
Shen_et_al_high_power_THz_radiation_PRL_2007}
and sensitive detectors%
\cite{Komiyama_et_al_THz_photon_detection_Nat_2000}
of THz radiation, will open up an extremely wide range
of possible THz applications, in particular,
in material diagnostics, probe nanoimaging,
biomedical applications etc. -- cf.\ Refs.\
\onlinecite{Ferguson_Zhang_THz_review_Nat_Mat_2002,
Tonouchi_THz_review_Nat_Phot_2007,%
Dragomans_THz_review_Progr_in_Quant_Electr_2004}.

Note that an alternative approach to the THz energy concentration using doped
semiconductor tapers has also been proposed.%
\cite{Nerkararyan_Hakhoumian_Babayan_Plasmonics_2008_THZ%
_Nanofocusing_in_Coax_Cone}
However, the required heavy doping of the semiconductors may cause fast
electron relaxation due to the collisions with the inflicted lattice
defects and bring about high losses. Therefore, in this Letter we will pursue the
adiabatic nanoconcentration of the THz radiation using metal/dielectric
structures.

Conventionally for THz and microwave regions, the metals are considered
as ideal which is equivalent to neglecting their skin depth
$l_s=\lambdabar_0/\mathrm{Re}\,\sqrt{-\varepsilon_m}$, where
$\varepsilon_m$ is the permittivity of the metal
(we take into accoint that in the THz region 
$\left|\varepsilon_m\right|\gg 1$), and
$\lambdabar_0=c/\omega$ is the reduced wavelength in vacuum.  It is true
that in the THz region  $l_s=30-60$ nm. i.e., $l_s\ll\lambda_0$.
However, as we show below in this Letter, it is the finite skin depth,
though as small as it is, that 
principally limits the ultimum localization size of the THz fields. For
larger waveguides, the THz wave energy is localized mostly in the vacuum
(dielectric) and its losses, which occur in the metal's skin layer, are
correspondingly small. The effective quality factor (or, figure of
merit) of the waveguide, which shows how many periods the wave can
propagate without significantly loosing its energy, can be estimated as
\begin{equation}
Q\sim 2a/l_s~,
\label{Q}
\end{equation}
where $a$ is the characteristic minimum size of the waveguide; this
estimate becomes a good approximation for a metal-dielectric-metal
planar waveguide [see below Eq.\ (\ref{slab_wavenum_1})].  When the
waveguide size reduces to become on the order of the skin depth,
$a\lesssim l_s $, the THz field is pushed into the metal, and the quality
factor reduces to $Q\lesssim 1$, which implies strong losses.
Qualitatively, this establishes the limit to the nanoconcentration: for
upper THz region $a\gtrsim l_s\approx 30$ nm, while for the 1 THz frequency
$a\gtrsim l_s\approx 60$ nm. These are the practical limits of the 
THz nanoconcentration for the noble metals (silver, gold, and platinum) 
and for aluminum.

If one pursues the goal of creating enhanced local fields in a small region,
but not necessarily to efficiently
transfer the THz energy from the far field to the near field, then the
apertureless SNOM approach, where a sharp metal or dielectric tip is irradiated by THz
radiation, can, in principle, achieve even higher resolution.%
\cite{Keilmann_J_Biol_Phys_2003_Review} However, the
efficiency of utilizing the THz energy of the source in this case
will be extremely low; the stray, far-field THz energy
may create a significant parasitic background.

\begin{figure}
\centering
\includegraphics[width=.48\textwidth]
{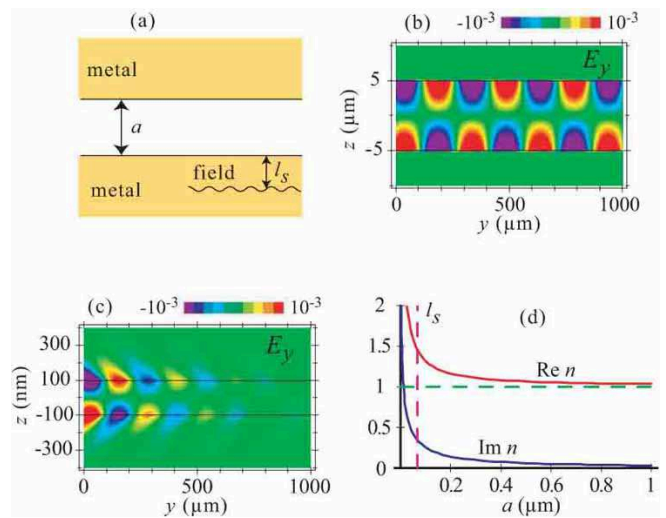}
\caption{\label{par_plate_guide.eps}
Geometry and properties of the THz TM mode in a parallel metal-slab waveguide.
(a) Schematic of the waveguide. The width of the
dielectric gap $a$ and the skin depth $l_s$ are indicated.
(b) An instantaneous distribution 
of the longitudinal electric field $E_y$  along the propagation
coordinate $y$ for $a=10~\mathrm{\mu m}$ and frequency 1 THz in a
silver-vacuum-silver waveguide. (c) The same as in panel (b) but for
$a=200$ nm. (d) Modal refraction index $n=k/k_0$ ($\mathrm{Re}\,n$ is
denoted by the red line and $\mathrm{Im}\,n$ by the blue line)
as a function of the waveguide width $a$. Dashed green line indicates
the value of
$n$ for the perfect conductor. Skin-depth value is shown by the vertical dashed
line.}
\end{figure}

Here and below in this Letter, we consider
examples of the THz adiabatic nanoconcentration
quantitatively, where the effect of the specific geometry
will become apparent.
Consider first a parallel plate waveguide that consists of a dielectric slab
of thickness $a$
with dielectric permittivity $\varepsilon_d$ sandwiched  between
two thick metal plates (with thickness of at least a few $l_s$, i.e., greater than
200 nm in practical terms) [see Fig.\ref{par_plate_guide.eps}(a)]. The
permittivity of the metal $\varepsilon_m$ in the THz region has a very
large ($\gtrsim 10^6$) imaginary part that defines the very small skin
depth $l_s\lesssim 100$ nm, which justifies the usual consideration of
the metals as perfect conductors.%
\cite{Ordal_fir_metal_dielectric_permittivities_ApplOpt_1983} However,
as we have already mentioned,
for our purposes of the THz nanoconcentration, we need to take into
account the field penetration into the metal, which makes the problem
plasmonic. In this case, the propagating modes of the system are SPPs,
which are TM modes characterized by the symmetry with respect to the
reflection in the center plane. We will orient the coordinate system
with its $z$ axis normal to the plane and the $y$ axis in the direction
of propagation. The symmetric (even) modes have even field components $H_x$ and
$E_z$ and odd $E_y$; the parity of the antisymmetric (odd) modes is opposite.

From plasmonics it is known that the even modes have a larger fraction
of their energy localized in the dielectric and the odd modes in the
metal. Therefore, the even modes have much smaller damping and are,
therefore, most suitable for the THz energy concentrations. The
dispersion relation for the even modes is given in the Methods section
as Eq.\ (\ref{slab_disp}).
This equation can be much simplified and solved in a closed analytical
form taking into account that we are interested in the subwavelength
focusing, i.e., $a\lesssim \lambdabar$, where
$\lambdabar=\lambdabar_0/\varepsilon_d$ is the reduced wavelength in the dielectric;
also, in the entire THz region $l_s\ll\lambdabar$. This shows that
there exists a small parameter in the problem $l_s a/\lambdabar^2\ll 1$
[see also Eq.\ (\ref{applicability})],
which allows one to solve analytically the dispersion relation
(\ref{slab_disp}) obtaining the modal refraction index $n=k/k_0$, where
$k$ is the THz wavevector, and $k_0=1/\lambdabar_0$,
\begin{equation}
n=\sqrt{\varepsilon_d}\left(1+\frac{l_s
(1+i)}{a}\right)^{1/2}\approx \sqrt{\varepsilon_d}
\left(1+i\frac{l_s}{2a}\right)~,
\label{slab_wavenum_1}
\end{equation}
where the approximate equality is valid for not too tight nanofocusing, i.e.,
for $l_s\ll a$. From this, we can obtain the quality factor of the
waveguide $Q=\mathrm{Re}\,n/\mathrm{Im}\,n=2a/l_s$, giving a quantitative
meaning to the estimate (\ref{Q}).

Plasmonic effects (i.e. those of the finite skin depth) are illustrated
in Fig.\ \ref{par_plate_guide.eps} for silver-vacuum-silver waveguide
and frequency of 1 THz. Panels (b) and (c) display the longitudinal
electric field $E_y$ obtained by the exact solution of the Maxwell
equations. Note that this field component is absent for the ideal
conductor; here it is relatively small: on
the order of $10^{-3}$ of the transverse field. 
Panel (b) illustrates the case of a relatively wide waveguide
($a=10~\mathrm{\mu m}$), where it is evident that
the electric field is localized mostly in the dielectric region of the
waveguide, and the extinction of the wave is small. In a sharp contrast,
for a nanoscopic waveguide ($a=200$ nm) in panel (c), the electric field
significantly penetrates the metal. In accord with our arguments, there
is a very significant extinction of the fields as they propagate; the
retardation effects are also evident: the lines of equal amplitude
are at an angle relative to the normal ($z$) direction. The dependence of the modal
refraction index on the thickness $a$ of the waveguide obtained from
Eq.\ (\ref{slab_wavenum_1}) is plotted in Fig.\
\ref{par_plate_guide.eps} (d). This index increases as $a$ becomes
comparable with the skin depth. While $\mathrm{Re}\,n$ and
$\mathrm{Im}\,n$ increase by the same absolute amount, the quality factor $Q$,
obviously, greatly decreases with decrease of $a$.
The mode described above can be used for broadband energy concentration
of THz waves.

\begin{figure}
\centering
\includegraphics[width=.48\textwidth]
{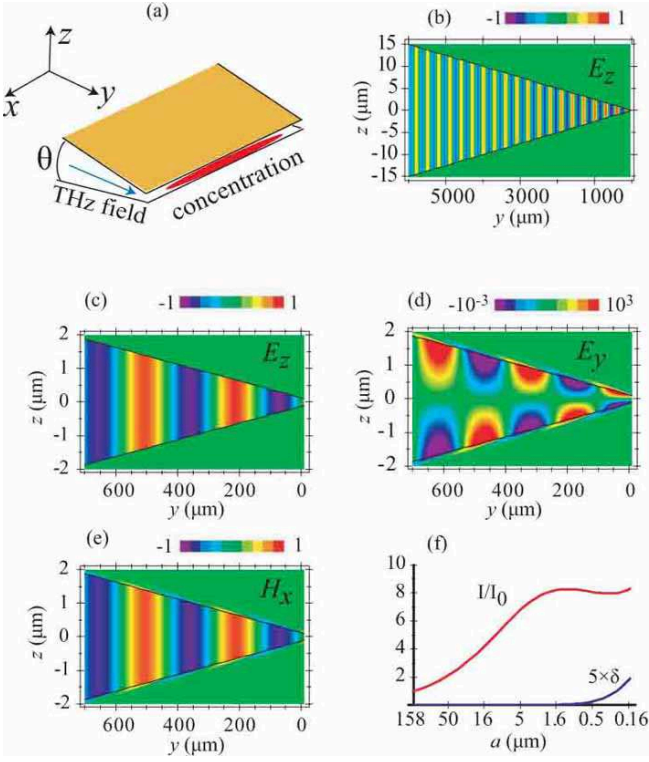}
\caption{\label{wedge.eps}
Adiabatic concentration of THz field energy in
a graded waveguide, where a dielectric wedge is surrounded by the thick
silver layer. (a) Schematic of energy
concentration, where $\theta$ is the wedge opening angle,
the arrow indicates the direction of propagation
of the THz wave, and the red highlights the area of the adiabatic
concentration. The orientation of the coordinate system is shown in the
inset. (b) An instantaneous distribution of the
transverse electric field $E_z$ of the THz wave
propagating and concentrating along the wedge waveguide for the last 6
mm of the propagation toward the edge.
Note the difference in scales in the $z$ and $y$ directions.
(c) An instantaneous spatial distribution of the transverse electric field $E_z$
close to the edge of the wedge, for the last $640~\mathrm{\mu m}$ of the
propagation.  (d) The same as (c) but for the
longitudinal (with respect to the propagation direction) component of
the field $E_y$. (e) The same as (c) but for the transverse component of
the magnetic field $H_x$. The units of these field components are arbitrary
but consistent between the panels. 
(f) Dependence of THz field intensity in the middle
of waveguide on the dielectric gap width $a$ (the red line). The blue
curve displays the dependence on $a$ of 
the adiabatic parameter $\delta$, 
scaled by a factor of 5. The values of $a$ indicated at the successive 
horizontal axis ticks differ by a factor of $10^{-1/2}$, i.e., by 5 dB.
}
\end{figure}

To introduce the THz nanoconcentration,
consider a metal-dielectric-metal waveguide that is slowly
(adiabatically) tapered off as a wedge, as illustrated in Fig.\
\ref{wedge.eps} (a). Because of the adiabatic change of the parameters,
a wave propagating in such a waveguide will adjust to it without
reflection or scattering, just as it takes place in nanoplasmonic
waveguides.\cite{Phys_Rev_Lett_93_2004_Tapered_Plasmonic_Waveguides} As
a result, propagating it will concentrate its energy, conforming to the
tapering of the waveguide. The corresponding solution can be obtained
from the Maxwell equations using the Wentzel-Kramers-Brillouin (WKB)
approximation, similarly to the nanoplasmonic case in the visible,%
\cite{Phys_Rev_Lett_93_2004_Tapered_Plasmonic_Waveguides} as described
in the Methods section. The WKB approximation is applicable under the 
conditions that
\begin{equation}
\delta=\left|d\left(\mathrm{Re}k^{-1}\right)/dy\right|\ll 1~,~~
\left| da/dy\right|\ll 1~,
\label{adiabaticity}
\end{equation}
where $\delta$ is the well-known adiabatic parameter describing how
slowly the modal wavelength changes on a distance of its own, and
$\left| da/dy\right|$ is a parameter describing how adiabatically the
transverse size of the confined mode changes along the propagation
coordinate.
 
In the WKB approximation, the behavior of the dominating
transverse field component $E_z$ as a function of the coordinate $y$
along the propagation direction is shown for 
the last $6~\mathrm{\mu m}$ of the propagation toward the edge
in Fig.\ \ref{wedge.eps} (b). There is a clearly seen
spatial concentration of the energy and increase of the field as the
wave is guided into the taper. The predicted behavior of the two
components of electric field and the magnetic field for the last micron
of the propagation is shown in panels (c)-(e). It apparently indicates
the adiabatic concentration, without an appreciable loss of the
intensity. The THz wave follows the waveguide up to the nanometric size.

The red line in Fig.\ \ref{wedge.eps} (f) indicates that the local
intensity $I$ as the function of the thickness $a$ of the waveguide for
$a<4~\mathrm{\mu m}$ increases significantly with $1/a$, in qualitative
accord with the behavior expected for the negligibly low losses. This
intensity reaches its maximum for $a=1.6~\mathrm{\mu m}$ and then starts
to decrease as the losses overcome the adiabatic concentration. At
smaller thicknesses, $a\lesssim 400$ nm, the intensity in Fig.\
\ref{wedge.eps} (f) starts to increase again, which is unphysical. The
reason is revealed by the behavior of the adiabatic parameter $\delta$
shown by the blue line: for $a\lesssim 400$ nm, $\delta$ becomes
relatively large (comparable with 1), i.e., the adiabaticity is
violated. This is due to the fact that the fraction of the THz field
energy propagating in the metal is dramatically increased for $a\lesssim
400$ nm due to the constricted transverse extension of the dielectric in
the waveguide. This causes a significant loss per wavelength $\lambda$,
leading to a rapid change of the wave vector $k$, breaking down the
adiabaticity. This constitutes a fundamental difference from the
nanoplasmonic adiabatic concentration in the optical region where the
adiabatic parameter is constant, and the adiabaticity holds everywhere
including the vicinity of the tip.%
\cite{Phys_Rev_Lett_93_2004_Tapered_Plasmonic_Waveguides}

\begin{figure}
\centering
\includegraphics[width=.48\textwidth]
{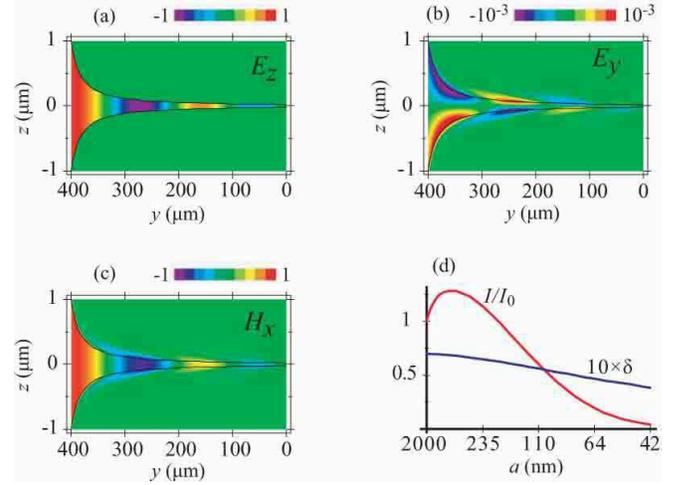}
\caption{\label{curved_wedge.eps}
Terahertz energy concentration in adiabatically tapered curved-wedge
waveguide.
(a) Instantaneous distribution of the transverse component of the THz
electric field $E_z$ (in the central plane $z=0$) as a function of the
coordinate $y$ along the propagation direction for the last
$400~\mathrm{\mu m}$ of the propagation. (b) The same as in panel (a)
but for the longitudinal electric field component $E_y$. (c) The same as
panel (a) but for the transverse magnetic field $H_x$. The units for the
fields are arbitrary but consistent between the panels. (d) The THz
field intensity $I$ (relative to the intensity $I_0$ at the entrance of
the waveguide) as a function of the dielectric gap thickness $a$ is
shown by the red line. The adiabatic parameter scaled by a factor of 10
as a function of $a$ is indicated by the blue line. The values of $a$
indicated at the horizontal axis ticks correspond to the values of $y$
at the ticks of panels (a)- (f).
}
\end{figure}

To provide for the optimum guiding of the THz wave and its concentration
on the nanoscale, the terminating (nanoscopic) part of the waveguide
should be tapered slower, in a funnel-like manner. That is, 
one needs to decrease
the grading $da/dy$ of the
waveguide near the edge in order to keep the adiabaticity parameter
$\delta=\left|d\left(\mathrm{Re}k^{-1}\right)/da\times da/dy\right|$
approximately constant and small enough to
prevent the back-reflection.
Because for the adiabatic grading (tapering), the derivative
$d\left(\mathrm{Re}k^{-1}\right)/da$
does not depend on the grading (it is the same as for the plane
waveguide) and is only a function of $a$, the equation $\delta=\delta(y)$ is
a differential equation for the shape of the waveguide that can be
easily integrated. This results in the dependence of the thickness $a$ on
the longitudinal coordinate $y$ determined by a simple integral
\begin{equation}
\mathrm{Re}\,n^{-1}(a)=k_0 \int\delta(y)dy~,
\label{grading}
\end{equation}
where $n(a)$ is the modal index defined in this case by
Eq.\ (\ref{slab_wavenum_1}), and $\delta(y)$ is the desired dependence
of the adiabatic parameter along the waveguide, which is an arbitrary
function of $y$ satisfying the adiabaticity conditions
(\ref{adiabaticity}).

The geometry of an adiabatically-tapered end of the silver/vacuum
waveguide found from Eq.\ (\ref{grading}) and satisfying Eq.\
(\ref{adiabaticity}) and the corresponding WKB solutions for the 1 THz
fields are shown in Fig.\ \ref{curved_wedge.eps} (a)-(c). The optimum
shape of the waveguide in this case is funnel-like, greatly elongated
toward the edge. The nanoconcentration of the field is evident on panels
(a)-(c), as well as its penetration into the metal for $a\lesssim 100$
nm. As these panels show quantitatively and the red curve on panel (d)
qualitatively, the field intensity reaches its maximum at $a\approx 300$
nm where it is enhanced with respect to the field at the entrance to the
funnel waveguide by a modest factor of 1.2. At the same time, the
adiabatic parameter $\delta$ decreases toward the tip from 0.07 to 0.05,
indicated the applicability of the WKB approximation everywhere. Note
that the this funnel-shaped wedge, indeed, continues the linearly-graded
wedge waveguide shown in Fig.\ \ref{wedge.eps}, which yields the
enhancement factor of $\approx 8$ at $a=2~\mathrm{\mu m}$. Sequentially,
these two waveguides provide the intensity enhancement by approximately
$\times 10$ while compressing the THz wave to the thickness of $a=300$
nm and the enhancement by a factor of 3 for $a=100$ nm.

Thus, true nanolocalization of THz radiation in one dimension (1d) is
possible. The minimum transverse size of this nanolocalization is
determined by the skin depth, as we have already discussed qualitatively
in the introductory part of this Letter. The obtained 1d beam of the
nanoconcentrated THz radiation may be used for different purposes, in
particular as a source for the diffraction elements including the
nanofocusing zone plates of the type introduced in Ref.\
\onlinecite{Merlin_Science_2007_Radiationless_Interference}. 

\begin{figure}
\centering
\includegraphics[width=.48\textwidth]
{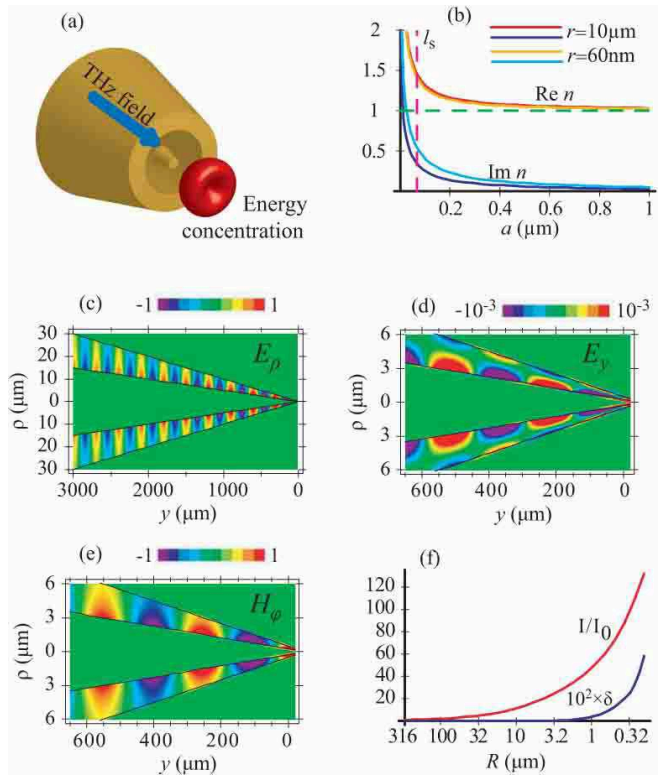}
\caption{\label{coax.eps}
Geometry, modal index of refraction, and THz energy concentration
in conically-tapered
metal-dielectric waveguide. (a) Schematic of geometry and energy
concentration. The central wire and the coax shell are shown along with
the schematic of the THz energy concentration.
(b) Dependence of modal refraction index $n$ in coaxial waveguide on the
dielectric gap width $a$ for two central wire radii: $r=10~\mathrm{\mu m}$
and $r=60$ nm. The color coding of the lines is indicated. The
dielectric in the gap is vacuum.
(c) Instantaneous distribution of the radial THz electric field amplitude $E_\rho$
in the cross
section of the coax for the last 3 mm of the propagation toward the
tip. The amplitude of the field is color coded by the bar at the top of the panel. 
(d) Instantaneous distribution of the longitudinal THz electric field
amplitude $E_y$ on
the coordinate $y$ for the last $620~\mathrm{\mu m}$ of the propagation.
(e) The same as (d) but for the transverse magnetic field $H_\varphi$.
The units of these field components are arbitrary
but consistent between the panels. 
(f) Dependence of THz field intensity in the middle
of waveguide gap on the waveguide outer radius $R=r+a$ is shown in red. The blue
curve displays the adiabatic parameter $\delta$ as a function of $R$, 
scaled by a factor of $10^2$. The values of $R$ indicated at the successive 
horizontal axis ticks differ by a factor of $10^{-1/2}$, i.e., by 5 dB.
}
\end{figure}

The two-dimensional (2d) concentration of the THz radiation can be
achieved by using an adiabatically-tapered conical coax waveguide, whose
geometry is illustrated in Fig.\ \ref{coax.eps} (a). The central metal
wire of radius $r$ is surrounded by a dielectric gap of the radial
thickness $a$, which is enclosed by a thick ($\sim 200$ nm or thicker)
outer metal shell. Both $r$ and $a$ are smooth functions of the
longitudinal coordinate $y$, which describes the tapering of the coax
toward the apex (tip) at $y=0$. The THz waves propagate from the wide
end of the coax toward the apex, adiabatically following the tapering.
In the spirit of WKB, for any particular $y$ the wave behavior for the
tapered coax is the same as for a cylindrical coax with the values of
$r$ and $a$ equal to the local values $r(y)$ and $a(y)$.

\begin{figure}
\centering
\includegraphics[width=.48\textwidth]
{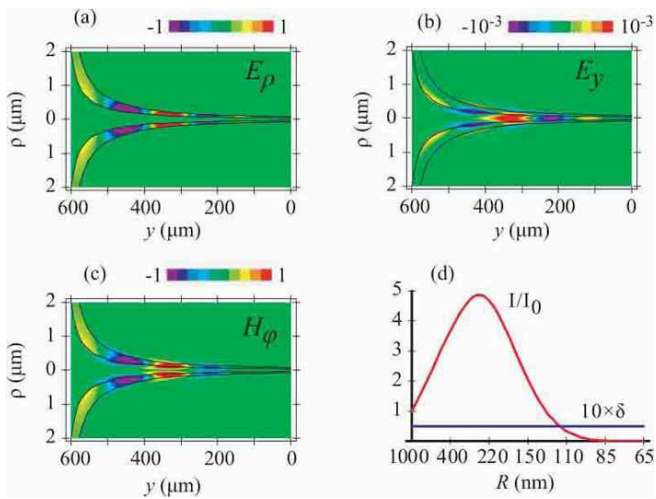}
\caption{\label{curved_coax.eps}
Adiabatic terahertz energy concentration in a self-similarly curved,
funnel-shaped
coaxial waveguide, where the metal is silver, and the dielectric in the
gap is vacuum. The dielectric gap is between the pair of the 
neighboring curved
lines, and the metal is everywhere else. (a) Instantaneous distribution
of the radial (transverse)
component $E_\rho$ of the electric field of the guided THz wave as a
function of the propagation coordinate along the wedge $y$ for the last 
$600~\mathrm{\mu m}$ of the propagation.  (b) The same
for the longitudinal electric field component $E_y$. (c) The same for
transverse magnetic field $H_\varphi$, whose lines form circles around the
central metal wire. The units of these field components are arbitrary
but consistent between the panels. (d) The THz intensity $I$ as a
function of the waveguide radius $R$, displayed relative to the intensity
$I_0$ at the beginning of the waveguide (red line). Adiabatic parameter
$\delta$ multiplied by a factor of 10 as a function of $R$ (blue line).
The values of the radius $R$ shown at the ticks correspond to those of
$y$ shown in panels (a)-(c).
}
\end{figure}

The dispersion relation for the coax waveguide that takes into account
the plasmonic effects (i.e., the penetration of radiation into the metal
and the concurrent losses) is obtained in the Methods section as Eq.\
(\ref{coax_refraction}). Calculated from this expression, the dependence
of the modal refractive index $n=k(a,r)/k_0$ on the dielectric gap $a$
is displayed in Fig.\ \ref{coax.eps} (b) for the frequency of 1 THz,
silver as a metal, and vacuum in the dielectric gap. The results are
shown for two values of the radius of the central wire:
$r=10~\mathrm{\mu m}$ and $r=60~\mathrm{nm}$. As one can see, the real
part of the modal index practically does not depend on $r$; it starts
growing when $a$ decreases. The imaginary part of the index $n$
increases when the central wire thickness $r$ decreases, but this
dependence is very weak. Both $\mathrm{Re}\,n$ and  $\mathrm{Im}\,n$ grow
dramatically for $r\lesssim l_s$. This is due to the penetration of the
THz field into the metal, i.e., it is a plasmonic effect.

The WKB solution for the radial field $E_\rho$ in the cross section of
this coax waveguide is shown for the last $3~\mathrm{\mu m}$ of the
propagation toward the tip in Fig.\ \ref{coax.eps} (c). The adiabatic
following and energy concentration are evident in this panel. The
penetration into the metal of the tangential (to the metal surface)
field components $E_y$ and $H_\varphi$ is noticeable in Figs.\
\ref{coax.eps} (d) and (e). The intensity $I$ of the THz field (relative
to the intensity $I_0$ at the entrance of the waveguide) as a function
of the waveguide outer radius $R=r+a$ is shown by the red line in Fig.\
\ref{coax.eps} (f). Dramatically, it shows the adiabatic
nanoconcentration and the intensity increase by more than two orders of
magnitude for the nanoconcentration from the waveguide radius
$R=300~\mathrm{\mu m}$, where the THz radiation can be focused, toward
$R=300$ nm. However, the dramatic increase of the adiabatic parameter
$\delta$ [plotted by the blue line in Fig.\ \ref{coax.eps} (f)] for
$R\lesssim 1~\mathrm{\mu m}$ shows that these results can only be
trusted for $R\gtrsim 1~\mathrm{\mu m}$. 

For the true 2d nanoconcentration of the THz radiation below this
micron-scale radius, similar to the 1d case of the wedge, to preserve
the adiabaticity, a funnel-like tapering is necessary. Generally, the
tapering of the central wire and that of the outer metal shell do not
need to be the same. However, we found that better results are obtained
when it is the case, i.e., the waveguide is tapered-off self-similarly.
In specific calculations, as everywhere in this Letter, we assume that
the metal of the waveguide is silver, the dielectric is vacuum, and the
frequency is 1 THz. Doing so, we have found the corresponding grading of
the waveguide using Eq.\ (\ref{grading}) and setting $\delta=0.05$,
which is small enough to satisfy the adiabaticity very well. In this
case, indeed, we have used the corresponding dispersion relation
(\ref{coax_refraction}). The obtained shape of the waveguide is a
strongly-elongated funnel, as shown in Fig.\ \ref{curved_coax.eps}
(a)-(c). These figures display the THz fields that we have calculated in
the WKB approximation for this waveguide. As one can see from these
figures, within the last half micron of the propagation, the electric
and magnetic fields of the THz wave efficiently follow the adiabatically
curved waveguide. The penetration into the metal of the tangential (to the
metal-dielectric interfaces) field components for $y< 400~\mathrm{\mu
m}$ is evident in panels (b) and (c). The longitudinal electric field
component $E_y$ is significantly localized in the central metal wire
[panel(b)], which is a plasmonic effect.

The dependence of the THz field intensity in the gap (relative to the
intensity $I_0$ at the entrance of this funnel) on the total radius of
the waveguide $R$ is shown in Fig.\ \ref{curved_coax.eps} (d) by the red
line. In this case, the adiabatic concentration is very efficient. The
intensity of the THz radiation increases by a factor of $\times 5$ when
it is compressed from the initial radius of $R=1~\mathrm{\mu m}$ to the
radius $R\approx 250$ nm. The penetration of the fields into the metal
for smaller values of the radius $R$ (tighter confinement) causes losses
that dominate over the effect of the concentration. Again, we remind
that this funnel waveguide is a continuation and termination for the
straigt cone that yields the field enhancement by $\times 50$ for
$R=1~\mathrm{\mu m}$ [see Fig.\ \ref{coax.eps} (f)]. Consecutively,
these two waveguides (the initial cone continued and terminated by the
funnel) are very efficient, adiabatically compressing the THz radiation
from the initial radius $R=300~\mathrm{\mu m}$ to the radius $R=250$ nm
increasing its intensity by a factor $\times 250$. Even for the final
radius $R=100$ nm, the total THz intensity is increased by a factor of
$\times 10$ (which is the products of factors $\times 50$ for the cone
part and $\times 0.2$ for the funnel. Thus, the optimally graded
plasmonic-metal 2d waveguide is very efficient in the concentration and
guidance of the THz fields with the transverse radius of confinement
$R\lesssim 100$ nm.

To discuss the results, we have shown that the THz radiation can be
concentrated to the $\sim 100$ nm transverse size in adiabatically
graded plasmonic (metal/dielectric) waveguides. In the optimum
adiabatically-graded, coaxial waveguide, which consists of the initial
cone terminated with a funnel, the radiation of a 1 THz frequency whose
wavelength is 300 $\mathrm{\mu m}$, can be compressed to a spot of 250
nm radius, where its intensity increases by a factor of $\times 250$.
Even in the case of the extreme compression to a spot of the 100 nm
radius, the THz intensity is enhanced by one order of magnitude with
respect to the initial intensity of the $300~\mathrm{\mu m}$ spot at the
entrance of the waveguide. The physical process that limits the extent
of this spatial concentration is the skin effect, i.e., penetration of
the radiation into the metal that causes the losses: the THz field
penetrates the depth of $l_s=30-60$ nm of the metal, which by the order
of magnitude determines the ultimum localization radius.

The THz nanoconcentration predicted in this Letter for optimally-graded
adiabatic plasmonic waveguides provides unique opportunities for THz
science and technology, of which we will mention below just a few. The
nanoconcentration of the THz radiation will provide the THz
ultramicroscopy with a THz source of unprecedented spatial resolution and
brightness. The increase of the THz intensity by two orders of
magnitude along with the novel high-power THz
sources\cite{Hebling_et_al_high_power_THz_sources_2008} would allow the
observation of a wide range of electronic and vibrational nonlinear
effects in metal, semiconductors, insulators, and molecules.

These nonlinear THz phenomena can be used to investigate behavior of
various materials in ultrastrong fields, for nonlinear spectroscopy
(including the multidimensional spectroscopy), and for monitoring and
detection of various environmental, biological, and chemical objects and
threats such as single bacterial spores and viruses. Such applications
will certainly be helped by very large absorption cross sections of
various materials in the THz region. A distinct and significant
advantage of the adiabatic nanofocusing is that the THz energy is mostly
concentrated in the hollow region of the waveguide, whose size can be
made comparable with the size of the objects of interest: in the range
from 1 micron to 70 nm, which is a typical range for bacteria and their
spores, and viruses. This will assure high sensitivity and low
background for the objects that are confined inside these waveguides.

Consider as a specific example the spectroscopy or detection of single
particles, such as, e.g., anthrax spores, in the air. A sample
containing the suspected nanoparticles in a gas, which can be air for
the frequencies in the transparency windows, can be pumped through a THz
waveguide, and the detection can be made for each particle in the gas
separately on the basis of the two-dimensional nonlinear THz spectra
that are expected to be highly informative for the detection and
elimination of the false-positive alarms. Likewise, many other
scientific, technological, environmental, and defense applications may
become possible.

\section[*]{Methods}
\label{Methods}

\subsection{Terahertz TM Wave in Finite-Conductivity
Parallel Plate Waveguide}

A parallel plate waveguide supports an even
TM mode with wavenumber $k$,
which satisfies the dispersion relation
\begin{equation}
\tanh{\left(\frac{\kappa_d a}{2}\right)}=
-\frac{\varepsilon_d\kappa_m}{\varepsilon_m\kappa_d}
\label{slab_disp}
\end{equation}
where $\kappa_d=\left(k^2-\varepsilon_d k_0^2\right)^{1/2}$,
$\kappa_m=\left(k^2-\varepsilon_m k_0^2\right)^{1/2}$.
In the terahertz range, $\varepsilon_m$ is mainly imaginary,
where $\mathrm{Im}\,\varepsilon_m\gg 1$.
Therefore, $\kappa_m\approx k_0\sqrt{-\varepsilon_m}=l_s^{-1} (1-i)$,
where $l_s=1/\text{Re}{\kappa_m}=\sqrt{2}/(k_0 \sqrt{|\varepsilon_m|})$
is the metal skin depth, which is on the order of tens of nanometers.
We also assume that $\kappa_d a\ll 1$, which is always the case
for the mode under consideration because either this mode is close to the
TEM mode where $k=k_0\varepsilon_d$, or the gap $a$ is thin enough.
This leads to a closed expression for the index of refraction
of the mode, which is Eq.\ (\ref{slab_wavenum_1}).
Using this, one can check that
$\kappa_d a\approx \left(\varepsilon_d a l_s/\lambdabar^2\right)^{1/2}$.
Consequently, the
applicability condition of the approximation used is
\begin{equation}
\left(\varepsilon_da l_s/\lambdabar^2\right)^{1/2}\ll 1~.
\label{applicability}
\end{equation}
This condition is satisfied for the realistic parameters of the problem.
For instance, for the frequency $f=1$ THz,
the skin depth for metals is $l_s\approx 60$ nm,
while reduced wavelength is $\lambdabar=75~ \mathrm{\mu m}$. The
condition (\ref{applicability}) is well satisfied for
$a\ll 100~\mathrm{\mu m}$, i.e., in the
entire range of interest to us.

\subsection{Terahertz TM Wave in Finite-Conductivity Coaxial Waveguide}

Consider a coaxial waveguide (coax) with the inner wire radius $r$
and the outer radius $R=r+a$, where $a$ is the dielectric gap width.
The characteristic relation for the TM modes of this waveguide has
has the following form
\begin{eqnarray}
&\left(\frac{I_0(\kappa_d r)}{I_0(\kappa_m r)}-
\xi\frac{I_1(\kappa_d r)}{I_1(\kappa_m r)}\right)
\left(\frac{K_0(\kappa_d R)}{K_0(\kappa_m R)}-
\xi\frac{K_1(\kappa_d R)}{K_1(\kappa_m R)}\right)=
\label{coax_characteristic}
\\\nonumber
&\left(\frac{K_0(\kappa_d r)}{I_0(\kappa_m r)}+
\xi\frac{K_1(\kappa_d r)}{I_1(\kappa_m r)}\right)
\left(\frac{I_0(\kappa_d R)}{K_0(\kappa_m R)}+
\xi\frac{I_1(\kappa_d R)}{K_1(\kappa_m R)}\right)~,
\end{eqnarray}
where $I_{\nu}(x)$ and $K_\nu (x)$ are modified Bessel functions, and
$\xi=\frac{\varepsilon_d \kappa_m}{\varepsilon_m \kappa_d}$.
This equation is quadratic with respect to $\xi$ and can be written in
the form $\alpha\xi^2+\beta\xi+\gamma=0$, where the coefficients
$\alpha$, $\beta$, and $\gamma$ can be easily found by comparison to
Eq.\ (\ref{coax_characteristic}) as combinations of the Bessel functions.
It can obviously be resolved for $\xi$ yielding
\begin{equation}
\frac{\beta\pm\sqrt{\beta^2-4\alpha\gamma}}{2\alpha}=
-\frac{\varepsilon_d \kappa_m}{\varepsilon_m \kappa_d}~.
\label{coax_disp}
\end{equation}

In the THz region, only the mode with the minus sign in Eq.\ (\ref{coax_disp})
propagates. It can be treated in a manner similar to the mode described by
Eq.\ (\ref{slab_disp}). The equation (\ref{coax_disp}) can be expanded over the small
parameter $\kappa_d a\ll 1$, and the explicit form of the modal refraction index
can be readily obtained as
\begin{eqnarray}
&n=\sqrt{\varepsilon_d}
\left(1+\left(\frac{I_0(\kappa_m r)}{I_1(\kappa_m r)}+
\frac{K_0(\kappa_m R)}{K_1(\kappa_m R)}\right)
\frac{l_s (1+i)}{2a}\right)^{1/2}~,~~
\label{coax_refraction}
\end{eqnarray}
where $\kappa_m=l_s^{-1}(1-i)$.
Similar to the wedge waveguide case, the applicability condition of this
solution is given by Eq.\ (\ref{applicability}).

\subsection{WKB Solution for the TM Wave in Graded Waveguide}

In the WKB approximation, a solution of the Maxwell equations can be represented
as a wave with amplitude and phase that are slowly
varying functions of $y$ on the scale of local wavelength.
The behavior in the transverse direction $z$ is the same as
for the non-graded system. The WKB solution is valid if
the adiabatic parameter is small
\begin{equation}
\delta=|\frac{d}{dy}\frac{1}{k(y)}|\ll 1~.~~
\label{adiabatic_parameter}
\end{equation}

In the WKB approximation,
the phase of the mode (eikonal) is given by an integral
\begin{equation}
\phi(y)=k_0 \int n(y) dy~,~~
\label{wedge_eikonal}
\end{equation}
where $n(y)$ is the local refraction index of the mode.
The behavior of wave amplitude as a function of the propagation
coordinate $y$ is found from the condition of flux conservation:
\begin{equation}
v_g(y) \int_{-\infty}^{\infty} W(y,z) dz = \text{const}~,~~
\label{flux_conservation}
\end{equation}
where $v_g(y)=\partial \omega/\partial k$ is the wave local group velocity,
and $W(y,z)$ is energy density in the mode.

This work was supported by grants from the Chemical Sciences,
Biosciences and Geosciences Division of the Office of Basic Energy
Sciences, Office of Science, U.S. Department of Energy, a grant
CHE-0507147 from NSF, and a grant from the US-Israel BSF. 
MIS is grateful to S.\ Gresillon for helpful remarks.

Correspondence and requests for materials
should be addressed to MIS~(email: mstockman@gsu.edu)


\end{document}